\documentclass[a4paper,fleqn,usenatbib]{mnras}

\usepackage[T1]{fontenc}
\usepackage{ae,aecompl}
\usepackage{relsize} 
\usepackage{color,soul}

\usepackage{graphicx}	
\usepackage{amsmath}	
\usepackage{amssymb}	
\usepackage{bm}

\title[Tidal Truncation of PPDs]{Protoplanetary Disc Response to Distant Tidal Encounters in Stellar Clusters}

\author[A.J. Winter et al.]{A. J. Winter,\thanks{E-mail: ajwinter@ast.cam.ac.uk} C. J. Clarke, G. Rosotti,  R. A. Booth
\\
Institute of Astronomy, Madingley Road, Cambridge CB3 0HA \\
}

\date{Accepted XXX. Received YYY; in original form ZZZ}

\pubyear{2017}

\begin{document}
\label{firstpage}
\pagerange{\pageref{firstpage}--\pageref{lastpage}}
\maketitle

\begin{abstract}
The majority of stars form in a clustered environment. This has an impact on the evolution of surrounding protoplanetary discs (PPDs) due to either photoevaporation or tidal truncation. Consequently, the development of planets depends on formation environment. Here we present the first thorough investigation of tidally induced angular momentum loss in PPDs in the distant regime, partly motivated by claims in the literature for the importance of distant encounters in disc evolution. We employ both theoretical predictions and dynamical/hydrodynamical simulations in 2D and 3D. Our theoretical analysis is based on that of \citet{Ost94} and leads us to conclude that in the limit that the closest approach distance $x_\mathrm{min} \gg r$, the radius of a particle ring, the fractional change in angular momentum scales as $(x_\mathrm{min}/r)^{-5}$. This asymptotic limit ensures that the cumulative effect of distant encounters is minor in terms of its influence on disc evolution. The angular momentum transfer is dominated by the $m=2$ Lindblad resonance for closer encounters and by the $m=1$, $\omega = 0$ Lindblad resonance at large  $x_\mathrm{min}/r$. We contextualise these results by comparing expected angular momentum loss for the outer edge of a PPD due to distant and close encounters. Contrary to the suggestions of previous works we do not find that distant encounters contribute significantly to angular momentum loss in PPDs. We define an upper limit for closest approach distance where interactions are significant as a function of arbitrary host to perturber mass ratio $M_2/M_1$.
\end{abstract}




\section{Introduction}

Star formation occurs preferentially in regions of enhanced gas density \citep{Lar81,Lada03}. It follows that stars generally form in clusters rather than in isolation, although the resulting properties of the cluster, including whether it remains bound after the local gas mass is expelled or accreted, is strongly dependent on the initial conditions \citep[e.g.][]{Lada03}. In such environments of enhanced stellar density, the number and significance of dynamical interactions between stars will increase accordingly. If sufficiently dense, such interactions are likely to have an impact on protoplanetary discs (PPDs). Indeed the differences between outer radius distributions for PPD populations in different stellar densities are statistically significant  for environments with a surface density $\bar{\Sigma}_\mathrm{c} > 10^{3.5}$~pc$^{-2}$ \citep{dJO12}. However, it remains unclear if this correlation is due to the tidal influence of cluster members, or additional physical effects such as external photoevaporation \citep[e.g.][]{Sca01, Cla07, Haw17}. 

While the focus of this work is the effect of encounters on PPDs, it is noted that the analytic work on perturbed binaries with non-eccentric orbits \citep{Heg96} is applicable to the test particle treatment of disc evolution, with associated induced eccentricities relevant to, for example, debris disc dynamics \citep{Ken01}. Historically, the theory of the tidal effects in this context has largely been divided into the influence of dynamical interactions between an existing binary and a perturbing star \cite[e.g.][]{Pre77, Heg93, Heg96}, and between a perturber and a single star with surrounding PPD \cite[e.g.][]{Gol78, Lub81, Ost94, Ogi02}. The implicit focus for a particular study in each case has either been on angular momentum transfer to the perturbing star \citep{Pre77} or to the unperturbed system (be it PPD or binary) \citep{Lub81}, or both \citep{Ost94}, depending on whether the authors are regarding mechanisms for stellar capture, tightening of a binary or induced accretion. In each case the relevant physical phenomena are similar, and require the contributions of various resonances between the natural frequencies of the unperturbed system and the trajectory of the perturber. 

Over the years, a range of computational approaches have been used to investigate the effect of star-disc encounters within clusters. This has included parameter exploration using test particles \citep{Cla93, Hal96, Pfa05,Pfa05a, Les11, Bha16}, statistical investigation of cluster dynamics combined with theoretical results \citep{Olc06, Vin16}, and hydrodynamical simulations of star cluster formation \citep{Bate12}. Most relevant  to our present investigation is the study of \citet{Ros14} who examined the effect of angular momentum loss from discs in the context of a cluster simulation with `live', viscously evolving discs. \citet{Ros14} found evidence that there is a significant range of encounter distances which, while not causing  mass loss from the disc, extract significant angular momentum from the outer disc and thus influence the growth of disc size. In our exploration of this effect, our re-analysis of the numerical data of \citet{Ros14} revealed no detectable effect of the  wider cluster environment on disc size \citep{Gio17}. It is  nevertheless possible that effects that would not be measurable over the duration of the simulation might prove to be important for real discs in clusters of sufficiently high density. Given the interest in this possibility that has been spurred by the \citet{Ros14} study, we here subject the suggested effect to detailed scrutiny.

The only detailed hydrodynamical study that has examined angular momentum transfer in non-pentrating disc encounters is that of \citet{Mun15}, which was focused on the evolution of the stellar components in a disc-disc interaction. {In that work the aim was not a comparison with the theoretical predictions for angular momentum transfer. It differs from this work in that both stellar components hosted a disc with a large relative mass, which complicates interpretation in the context of disc evolution, especially as many of the models involved strong disc-disc interactions. Our focus in this paper is to produce a robust general expression for encounter induced angular momentum loss within the disc in the linear, low disc mass regime as a function of orientation, stellar mass ratio and closest approach distance.}

We adopt the following approach in examining the angular momentum transfer in non-penetrating star-disc encounters. First we develop the linearised equations for the transfer of angular momentum between a ring of particles and a stellar pair undergoing a parabolic flyby. Our approach bears similarities to that of \citet{Ost94} but we have re-derived relevant expressions for several reasons. Most importantly, that study was concerned with the case of young massive discs which might be relevant to the formation of binary stars by capture. We, by contrast, are concerned with the progressive influence of multiple encounters throughout the pre-main sequence period and therefore need to treat the case that the disc mass is small compared with the mass of the stars. As we shall see, some of the expressions from \citet{Ost94} should clearly not be applied to the case of low disc mass since they predict an infinite change in specific energy and angular momentum in the test particle limit. Further, Ostriker does not present explicit expressions for arbitrary stellar mass ratios and relative phase between pericentre and the line of nodes of the disc and stellar orbits, nor for the dependence of the angular momentum transfer on disc surface density profile. 

 We test the linearised expressions we derive by comparison with numerical integration of the response of a ring of test particles to a parabolic perturber. Smoothed particle hydrodynamic (SPH) simulations are then used to reproduce this calculation in a disc including pressure and viscosity forces, which we compare to a disc reconstructed with an appropriate surface density profile from the test particle ring results. We find excellent agreement between the linearised expressions, the test particle calculations and the SPH simulation results, thus validating the use of SPH to correctly model star-disc encounters in the linear regime. We will nevertheless find that the asymptotic fall-off of the angular momentum transfer with pericentre radius implies that the cumulative effect of distant encounters is small.

The rest of this work is organised into the following sections. Section \ref{sec:method} reviews the relevant equations and our modelling techniques. Our numerical results are laid out in Section \ref{sec:results}, and these are discussed in the context of a stellar cluster in Section \ref{sec:discuss}. Our conclusions are summarised in Section \ref{sec:conclusion}.

\section{Theory and Method}
\label{sec:method}

\subsection{Linearised Equations}
\label{sec:lineqs}
In order to compare theoretical angular momentum transport in discs to our models, we modify the relevant equations in \citet{Ost94}. These results are a first order approximation for the change in the magnitude of angular momentum in a PPD with original angular momentum vector $L_z \hat{\bm{e}}_z$ during an encounter for which the minimal separation between stellar components $|\bm{x}(t)|_{\mathrm{min}} = x_{\mathrm{min}}$ occurs at $t=0$. In this case, for $\Delta L \ll L$, it is easy to show that
\begin{equation}
\label{eq:linapprox}
\frac{\Delta L}{L} = \frac{|\bm{L}|_{t\rightarrow \infty} - |\bm{L}|_{t\rightarrow -\infty}}{|\bm{L}|_{t\rightarrow -\infty}} = \frac{\Delta L_z}{L_z}+ \mathcal{O}(\Delta L_{\perp}^2/L_z^2)
\end{equation} where $L_{\perp}$ is the change of angular momentum perpendicular to $\bm{e}_z$. Hence the linearised equations are concerned with the change parallel to the original angular momentum vector of the disc.

The way that \citet{Ost94} calculates this is by first decomposing  $\bm{f}_{\mathrm{ext}}$, the external force exerted by the star per fluid element in the disc, into spherical harmonics. Angular momentum change per unit time per fluid element is then found by taking the $\hat{\bm{e}}_z$ projection of the cross product with the position vector of the fluid element relative to the central star. The angular momentum change of the fluid element is then obtained by integrating the torque over time. For a fluid element that remained on  a circular orbit throughout the interaction, the total angular momentum change associated with the interaction would be zero, as discussed by \citet{Heg96} in the context of perturbations to a non-eccentric binary. Angular momentum transfer is associated with the torque acting on the fluid element's {\it perturbed} trajectory. This is evaluated by considering the temporal Fourier decomposition of the element's perturbations and the  interaction between each Fourier component and the corresponding Fourier component of the external force. For each fluid element, disturbances are excited at frequencies which correspond to resonances with the forcing frequency of the perturbing star \citep{Gol78}. For a central potential $\bm{\Phi}_0$, the natural frequencies within the disc are the circular, epicyclic and vertical angular frequencies, for which, at radius $r_0$ in an unperturbed disc, are

$$
\Omega_0^2 \equiv  \frac 1 {r_0} \frac{\partial \Phi_0}{\partial r_0} \,; \quad \kappa_0^2 \equiv  \frac 1 {r_0^3} \frac{\partial \left( r_0^4 \Omega_0^2\right)}{\partial r_0}  \, ; \quad \chi_0^2 \equiv  \left. \frac{\partial^2 \Phi_0}{\partial z^2}\right|_{r_0;\, z=0} 
$$ respectively. For a disc of negligible mass, this means that
$$
\Omega_0^2 = \kappa_0^2 = \chi_0^2 = \frac{GM_1}{r_0^3}
$$ where $M_1$ is the mass of the disc hosting star. These frequencies are associated with corotation, Lindblad and vertical resonances respectively. For a given azimuthal wavenumber $m$, the equations
$$
m \Omega_0 - \omega =0 \,; \quad m \Omega_0 - \omega = \pm \kappa_0 \,; \quad m \Omega_0 - \omega = \pm \chi_0
$$ can be solved for a corresponding radius $r_0$ (we henceforth drop the subscript) at each forcing frequency $\omega$. The positive or negative Lindblad and vertical frequencies correspond to inner and outer resonances respectively. Note that, while angular momentum transfer within the disc is associated exclusively with radii in resonance with the forcing frequency $\omega$, $\bm{f}_\mathrm{ext}$ is Fourier decomposed such that every location in the disc is always in resonance with some component of the forcing potential since it has a continuum spectrum.

The evaluation of the angular momentum transfer from these resonances is discussed more fully in Appendix \ref{sec:reqns}. We note that the results of \citet{Ost94} include an inner vertical resonance (IVR) term which does not in fact contribute to angular momentum transfer to first order \citep[see Appendix \ref{sec:reqns} and][]{Lub81}. This means that the dominant resonances are the inner Lindblad resonances (ILRs), resulting in a steeper asymptotic power law of $x_\mathrm{min}/r$ than if the IVR contributed to first order. We further note that, as previously discussed, the focus of \citet{Ost94} was the regime of high (order unity) disc to star mass ratio and hence was developed to address this limit. This formulation is clearly not to be used in the limit of low disc mass since it predicts an infinite change in {\it specific} energy and angular momentum in the test particle limit \citep[cf. Equation 2.48 in][]{Ost94}. Instead our calculation leads to an expression, Equation \ref{eq:NLR}, equivalent to the test particle result quoted (but not employed) in Equation [2.50] of that study.

For all values of $x_{\mathrm{min}}/r$ there are just two dominant contributions to angular momentum loss in the disc. They are the $m=2$ ILR for close encounters ($x_{\mathrm{min}}/r \lesssim 6$ for equal mass stellar components) or the $m=1$ ILR for larger $x_{\mathrm{min}}/r$, where the latter corresponds the limit of small forcing frequency $\omega=0$, as discussed in Appendix \ref{sec:reqns}. These contributions are evaluated in Equations \ref{eq:ILR_an} and \ref{eq:NLR} respectively.

\subsection{Ring of Test Particles}
\label{sec:ringmeth}

To assess the effect of a stellar encounter at a single radius within a disc, a test particle calculation is applied. We use the general Bulirsch-Stoer algorithm of the \textsc{Mercury} orbital integrator for solar-system dynamics \citep{Cha99}.

We set up a ring of $200$ test particles at $r=1$~au from a central star with mass $M_1= 1M_\odot$. A second star of mass $M_2$ is placed on a parabolic trajectory at a time $200$ test particle orbits prior to closest approach, and integrated for the same time subsequent to that approach. The system is further defined by two angles: the angle between the direction of pericentre and the line of intersection of the disc and the orbital plane, $\alpha$, and the angle between the angular momentum vector of the disc and that of the orbit, $\beta$. Note that the system is scale free in that the only pertinent quantities are the ratio of the closest approach to test particle ring radius $x_{\mathrm{min}}/r$ and the mass ratio between the stellar components $M_2/M_1$. 

The specific angular momentum for each particle is then compared to the corresponding particle in a ring which remains unperturbed over the same period, and the average angular momentum loss over all the particles represents the total loss for the ring. {For encounters for which $x_\mathrm{min}/r \lesssim 2$ some particles become unbound from the initial star. For an unbound particle (i.e. one with a post-encounter eccentricity $e>1$) we remove the angular momentum of that particle from the disc. This choice does not influence the majority of results, especially in the distant regime of interest. If we chose to ignore unbound particles and average only over those remaining, then at $x_\mathrm{min}/r =2$ this only changes the recorded angular momentum loss by $\sim 10\%$ for prograde encounters, and less when $\beta \neq 0^\circ$ for which encounters are less destructive.}

\subsection{Hydrodynamic Modelling}
\subsubsection{Numerical Method}
\label{sec:nummethod}
The smooth particle hydrodynamics (SPH) code \textsc{Gandalf} \citep{Hub16} is used to simulate a star-disc interaction including hydrodynamic forces.  The encounters under consideration are non-penetrative, and therefore we do not expect strong shocks. For this reason we apply the $\alpha$-viscosity formulation of \citet{Mon97} to model the viscous redistribution of angular momentum throughout the disc with a lower value of $\alpha_\mathrm{AV}=0.1$ so that viscous evolution is slow. Particles are integrated using the leapfrog kick-drift-kick integration method with a cubic spline kernel. We do not include the self gravity of the SPH particles, or the gravitational effects of the discs on the stars. However, any accreted particles contribute to the mass of the associated sink particle. The smoothing length for both sink particles is defined to be half the inner radius of the disc, $R_\mathrm{in}/2$.

In order to investigate the possible influence of the finite thickness of the disc upon the angular momentum transfer, we compare our 3D calculations with equivalent 2D SPH calculations for the cases of co-planar prograde and retrograde encounters. The resolution at 3D is fixed at $10^6$ particles. In order to reproduce the calculations in 2D, {the variable smoothing length $h_i$ for particles at the same radius $r_i$ in the disc needs to be equivalent.} In other words:
$$
    h_i = \eta \left( \frac{m_{3\mathrm{D}}}{\rho_i} \right)^{1/3} = \eta \left( \frac{m_{2\mathrm{D}_i}}{\Sigma_{\mathrm{2D}}(r_i)} \right)^{1/2}
$$ where $m_{3\mathrm{D}}$ is the (constant) mass of each particle in 3D, and  $m_{2\mathrm{D}_i}$ is mass of the equivalent particle in 2D, which is position dependent.The surface density in 2D $\Sigma_{2\mathrm{D}}$ for a given particle at radius $r_i$ must also be equal to the surface density in 3D such that
$$
\Sigma_{2\mathrm{D}}(r) = \Sigma_{3\mathrm{D}}(r) = \Sigma(r) 
$$ for a prescribed surface density profile $\Sigma$. Hence the mass $m_i$ of a particle at a given radius $r_i$ is
$$
m_{2\mathrm{D}_i} = \left\{\sqrt{2\pi\Sigma(r_i)}H(r_i)m_{3\mathrm{D}}\right\}^{2/3}
$$ where $H(r)$ is the scale height of the disc, which is defined in Section \ref{sec:SPHICs}, along with our chosen surface density profile. For our physical parameters we find a corresponding 2D resolution of $\sim 10^5$ particles to compare with the 3D version with $10^6$ particles.

\subsubsection{Physical ICs}
\label{sec:SPHICs}

We choose disc parameters consistent with those of \citet{Ros14}, in which hydrodynamic disc evolution is studied in a stellar cluster of $100$ stars, so that a comparison can be drawn with their results. These conditions are described as follows.

The surface density follows a truncated power law 
\begin{equation}
\label{eq:plsd}
\Sigma(r) = \Sigma_0 \left(\frac{r}{r_0} \right)^{-p}
\end{equation} for $p=3/2$, where $r_0$ and $\Sigma_0$ are the scale radius and surface density scale respectively, and are such that the total mass of the disc is $0.05M_\odot$, although as discussed in Section \ref{sec:nummethod}, this does not modify the overall gravitational potential. \citet{Ros14} choose a range of disc radii, but here we choose the model referred to as R10 in that study. This model has an outer disc radius $R_\mathrm{out} = 10$~au and inner radius of $R_\mathrm{in} = 2$~au around a star of mass $1\, M_\odot$. 

A locally isothermal equation of state is chosen, with a temperature varying with radial distance from the star such that the temperature is
$$
T(r) = \mathrm{max} \left[  T_0 \left( \frac{r}{r_0} \right)^{-q}, \, 20\mathrm{ K} \right]
$$  where $q=3/2$. The height of the disc $H=c_s/\Omega$, where $c_s$ is the sound speed and $\Omega$ is the Keplerian frequency, is chosen so that $H/r$ is 0.05 at the inner radius. This implies that $T_0=20$~K for $r_0=14$~au.

{The disc is evolved for $\sim 12$ orbits at the (viscously evolving) outer radius before and after the stars reach the closest approach distance. This is sufficient for the disc to `relax' prior to closest approach but short enough so that the viscous evolution has not significantly altered the surface density profile. As has been found in previous studies} \citep[e.g][]{Hal96} {we find that angular momentum transfer occurs when the phase of the perturbing star is close to pericentre, and for all our SPH results this corresponds to a time span which is $\lesssim \left[\Omega(R_\mathrm{out})\right]^{-1}$, the orbital period at the outer edge of the disc.}

\section{Numerical Results}
\label{sec:results}
\subsection{Perturbed Ring}
\label{sec:ring_results}

\begin{figure*}
	\includegraphics[width=\textwidth]{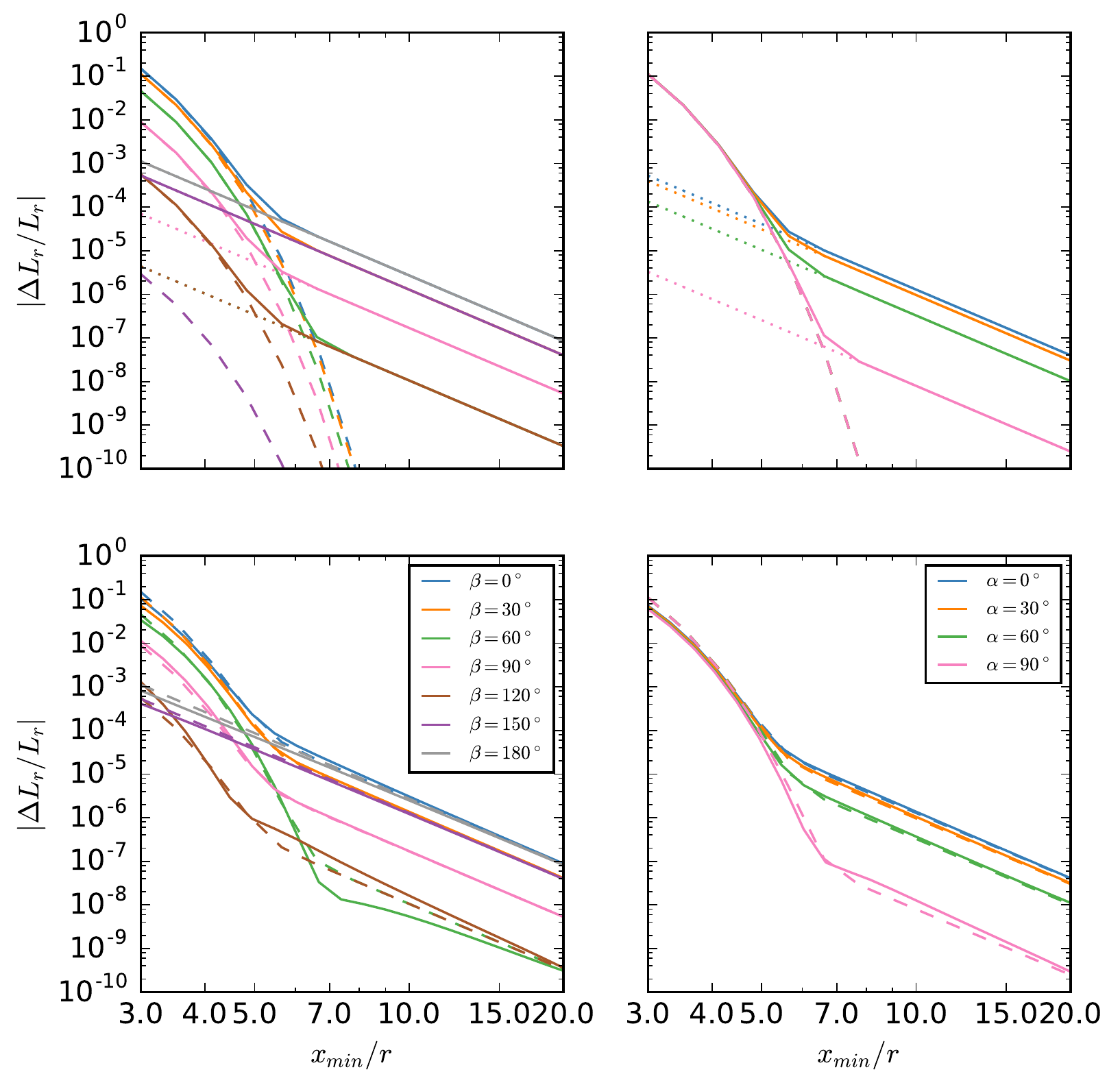}
    \caption{Change of angular momentum for a ring around a central star due to a parabolic encounter between stars of equal mass. The results on the left are for $\alpha=0^\circ$ and various $\beta$ values, while those on the right are for $\beta=30^\circ$ and varying $\alpha$. In the top panels are the results of evaluating Equations \ref{eq:NLR} (dotted lines) and \ref{eq:ILR_an} (dashed lines) for the linear approximation of the fractional angular momentum change, with the solid line showing the sum of the two components. In the bottom plots, the dashed lines are the theoretical results, while the solid lines are the results for a ring of test particles.}
    \label{fig:dLringall}
\end{figure*}

\begin{figure*}
	\includegraphics[width=\textwidth]{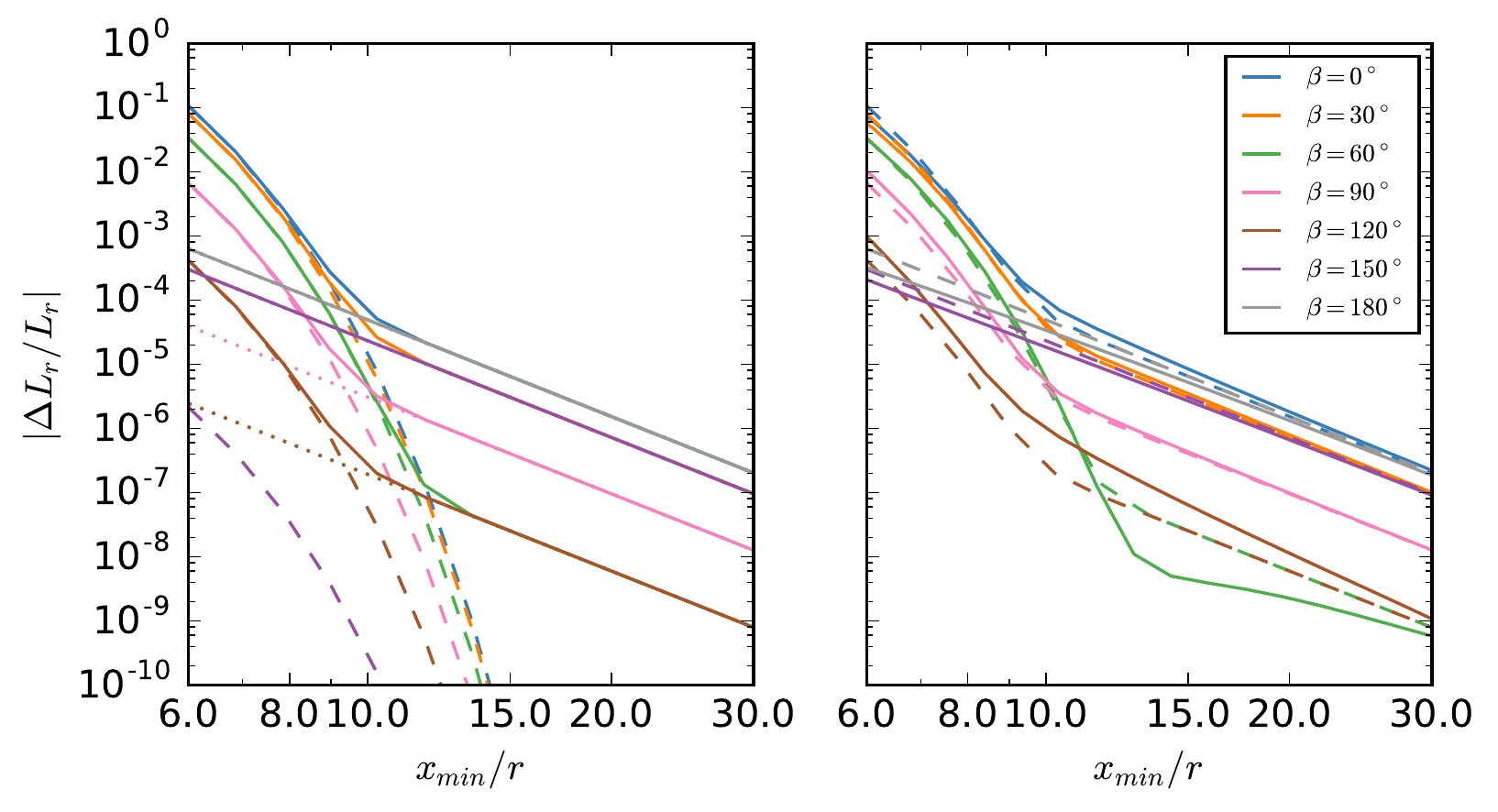}
    \caption{Fractional angular momentum loss for a massless ring around a central star of mass $M_1$ when perturbed by a star of mass $M_2$ such that $M_2/M_1=10$. Results are shown for $\alpha=0^\circ$ and varying $\beta$. In the left panel, the evaluation of Equations \ref{eq:NLR} and \ref{eq:ILR_an} are shown in dotted and dashed lines respectively, while the sum is shown as a solid line. In the right panel the solid lines are for a ring of test particles, with the dashed lines being the theoretical counterparts.}
    \label{fig:dLringmrat10}
\end{figure*}

The sum of the contributions to $\Delta L/L$ from the numerical integrations in Equations \ref{eq:ILR_an} ($m=2$ ILR, exponential term) and \ref{eq:NLR} ($m=1$ $\omega=0$ ILR, power law contribution) and the results for a ring of test particles are plotted in Figure \ref{fig:dLringall} for various orientations. For nearly all regions of the parameter space these results are within order unity of the theoretical counterparts; certainly this is the case for all results for which $\Delta L/L >10^{-5}$. There is some deviation in the results for which the trajectory is highly inclined, in particular for the $\beta \approx 60^\circ, 120^\circ$ results. This is likely to be because the linear estimate for $\Delta L/L$ is made assuming the change in angular momentum is dominated by the change in the initial direction of $\bm{L}$, which is $\Delta L_z$ (see Appendix \ref{sec:reqns}), whereas this is not strictly true for encounters with a periastron close to perpendicular to the disc plane. However, the regions of parameter space for which the results deviate significantly are those for which angular momentum loss is negligible. This is still true when the mass ratio between the perturbing and central star $M_2/M_1$ is increased, as shown in Figure \ref{fig:dLringmrat10}. For this reason we do not investigate the deviation further in this study.

In the limit of large closest approach distances $x_{\min}$, the angular momentum loss for a ring at radius $r$ scales as $(x_{\min}/r)^{-5}$, which corresponds to the the contribution of the $m=1$, $\omega=0$ ILR, as approximated by Equation \ref{eq:NLR}. At closer periastron distances, the exponential component from the exact resonances in Equation \ref{eq:ILR_an} dominate for prograde and inclined trajectories. 

\begin{figure}
	\includegraphics[width=\columnwidth]{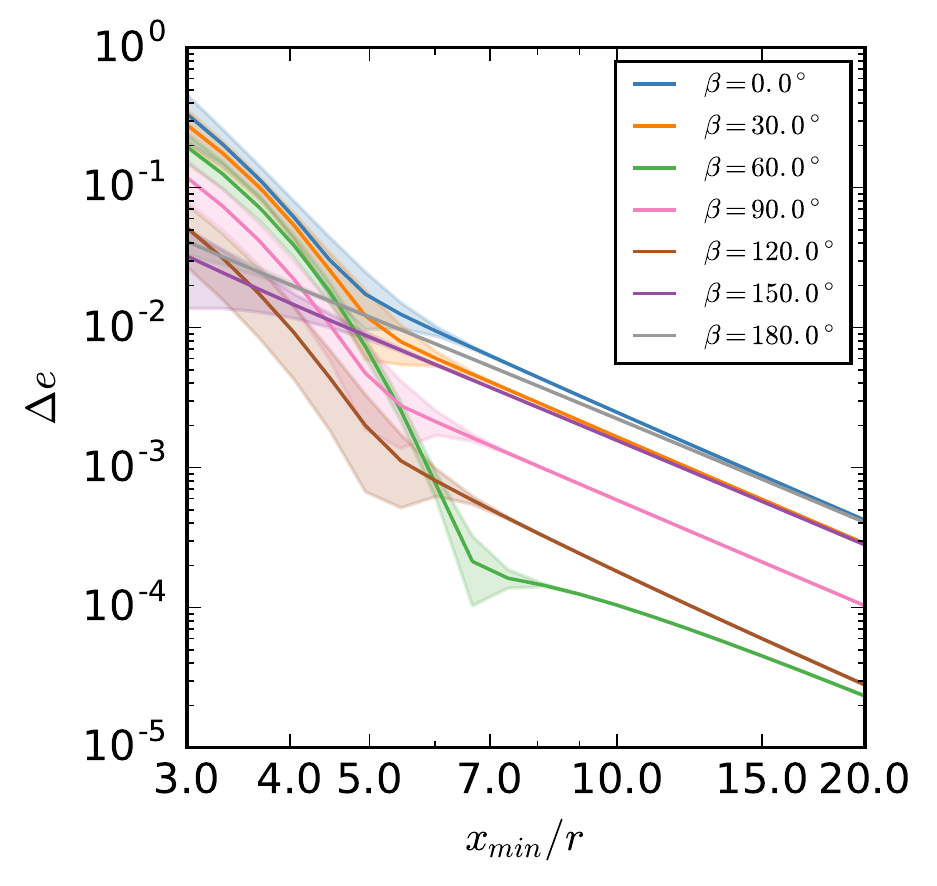}
    \caption{Change in eccentricity $\Delta e$ of a ring of initially non-eccentric test particles at radius $r$ induced by a parabolic stellar encounter with closest approach $x_\mathrm{min}$ in which both stellar components have equal mass. The shaded region around each line indicates the standard deviation of the induced particle eccentricities around each result.}
    \label{fig:dering}
\end{figure}

The eccentricity perturbations induced by the tidal disruption are shown in Figure \ref{fig:dering}. As angular momentum is proportional to $\sqrt{1-e^2}$, then $\Delta L \propto \Delta e^2$ for $\Delta e \ll 1$ as the particles are initially on a circular ($e=0$) orbit. Indeed the change in eccentricity scales with $(x_{\mathrm{min}}/r)^{-5/2}$ in the limit of distant encounters, which is in agreement with the results of \citet{Heg96} for an initially non-eccentric binary. A quantitative comparison with the results of \citet{Ost94} is made in the following Section \ref{sec:ringint}.

The ratio of the mass of the perturbing star $M_2$ to that of the host star $M_1$ also influences the angular momentum loss. From Equation \ref{eq:NLR} we expect the asymptotic dependence $x_\mathrm{min} \gg r$ to be $\Delta L/L \propto M_2/M_1$ for $M_2 \gg M_1$. In the close encounter regime however, the exponential component (Equation \ref{eq:ILR_an}) is expected to dominate out to greater $x_\mathrm{min}/r$. This is confirmed in Figure \ref{fig:dLringmrat10} for both test particle and theoretical calculations with $M_2/M_1=10$. The loss at $x_\mathrm{min}/r \sim 6$ is found to be more than two order of magnitudes larger in this case than for $M_2/M_1=1$.

\subsection{Perturbed Disc}

\subsubsection{Ring Integration Results}
\label{sec:ringint}

\begin{figure}
	\includegraphics[width=\columnwidth]{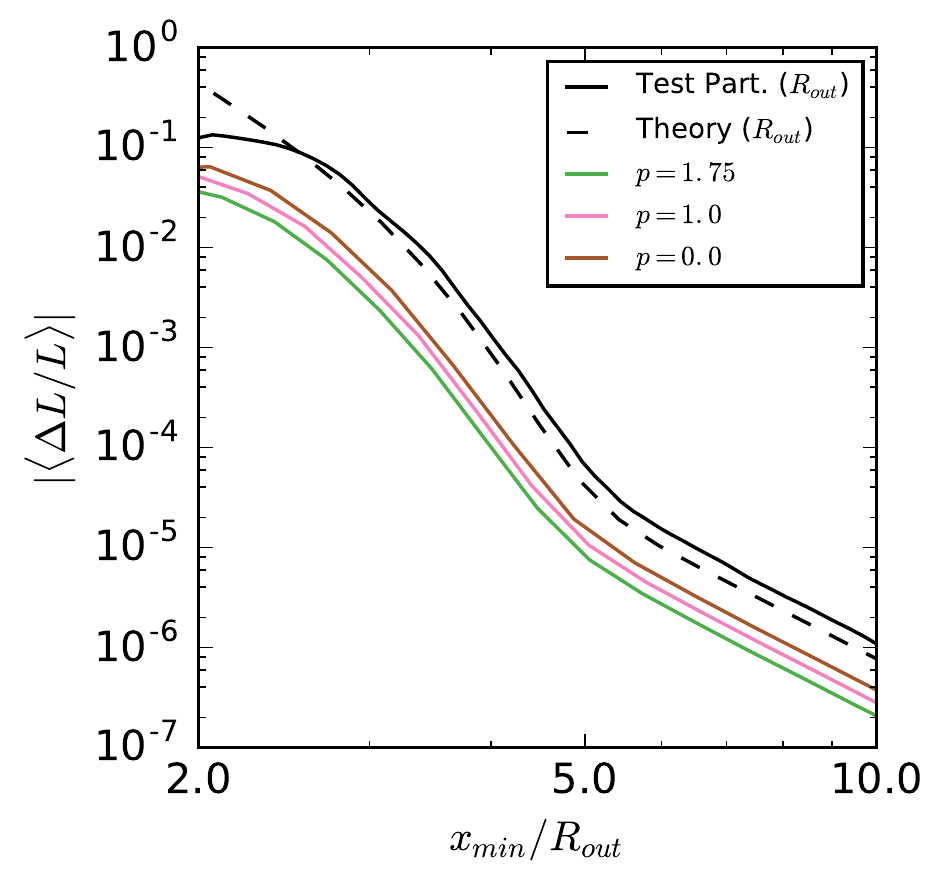}
    \caption{Angle averaged angular momentum transfer due to an equal mass perturber for discs contructed from test particle rings with surface densities following various powerlaws $\propto  r^{-p}$ and $R_\mathrm{out}/R_\mathrm{in}=5$. The $p=1.75$ and $p=1$ results are comparable to the results in Figure 5 of \citet{Ost94}. The dashed line represents the theoretical angular momentum loss for a ring with radius $R_\mathrm{out}$, and this is compared with the ring of test particle case (solid black line).}
    \label{fig:dL_rdisc_angavg}
\end{figure}

In order to draw useful conclusions regarding cluster dynamics, we present angle-averaged results. The angle averaging is simply the integral over the solid angles such that
$$
\left\langle \frac{\Delta L_\mathrm{r}}{L_\mathrm{r}} \right\rangle = \frac 1 {4\pi} \int_0^{2\pi} \mathrm{d} \alpha \int_0^{\pi}\mathrm{d}\beta \, \sin\beta\frac{\Delta L_\mathrm{r}}{L_\mathrm{r}}(\alpha, \beta) 
$$ evaluated at any given $x_\mathrm{min}/r$. 

We also present, for comparison with \citet{Ost94}, the results applied to a synthetic disc that is composed of a suitably weighted ensemble of particle rings which correspond to the same surface density profiles. The fractional change of angular momentum is 

\begin{equation}
\label{eq:reconst}
 \left. \frac{\Delta L_{\mathrm{d}}}{L_\mathrm{d}}\right|_{x_{\mathrm{min}}}  \! \! \! \! \! =  \frac{\int_{R_{\mathrm{in}}}^{R_{\mathrm{out}}} \! \! \mathrm{d}r \, r \Sigma(r) \left\langle \Delta L_\mathrm{r} \left(x_\mathrm{min}/r \right)  \right\rangle } {\int_{R_{\mathrm{in}}}^{R_{\mathrm{out}}} \! \! \mathrm{d}r \, r \Sigma(r)   L_\mathrm{r}(r) }
\end{equation} for a disc with an arbitrary surface density profile $\Sigma$. For a low mass disc composed of test particles, orbits are Keplerian such that $L_\mathrm{r} \propto r^{1/2}$.

The angle averaged results for a ring of particles, and discs as described in Equation \ref{eq:reconst} applied to power law surface densities (Equation \ref{eq:plsd}) are presented in Figure \ref{fig:dL_rdisc_angavg} for various values of $p$, truncated at a given outer radius $R_\mathrm{out}$. We note that practically, because we do not have numerical test particle results for $x_\mathrm{min}/r \rightarrow \infty$, contributions have to be truncated for small radial extents $r$ within the disc, and we choose $R_\mathrm{out}/R_\mathrm{in} = 5$. The inner disc contributions would be negligible, and the results in Figure \ref{fig:dL_rdisc_angavg} are dominated by the surface density at the outer radius. This is expected given the strong dependence on $x_\mathrm{min}/r$ for the ring results.

Our results agree with those of \citet[][see Figure 5 therein]{Ost94} in the exponential ILR regime described by Equation \ref{eq:ILR_an}, as expected. The asymptotic slope differs slightly however, as in that study the vertical resonance was considered dominant, such that $\Delta L_\mathrm{d} \propto (x_\mathrm{min}/r) ^{-4.5}$. By contrast we find a power law index of $-5$ as predicted in Equation \ref{eq:NLR}. Quantitatively our results are an order of magnitude lower at $x_\mathrm{min}/R_\mathrm{out} = 5$ and a factor $\sim 30$ lower at $x_\mathrm{min}/R_\mathrm{out} =10$. However, since $\Delta L/L$ is small in this region, these differences are not of practical significance.

\subsubsection{SPH Disc Results}
\label{sec:sphresults}

\begin{figure*}
	\includegraphics[width=\textwidth]{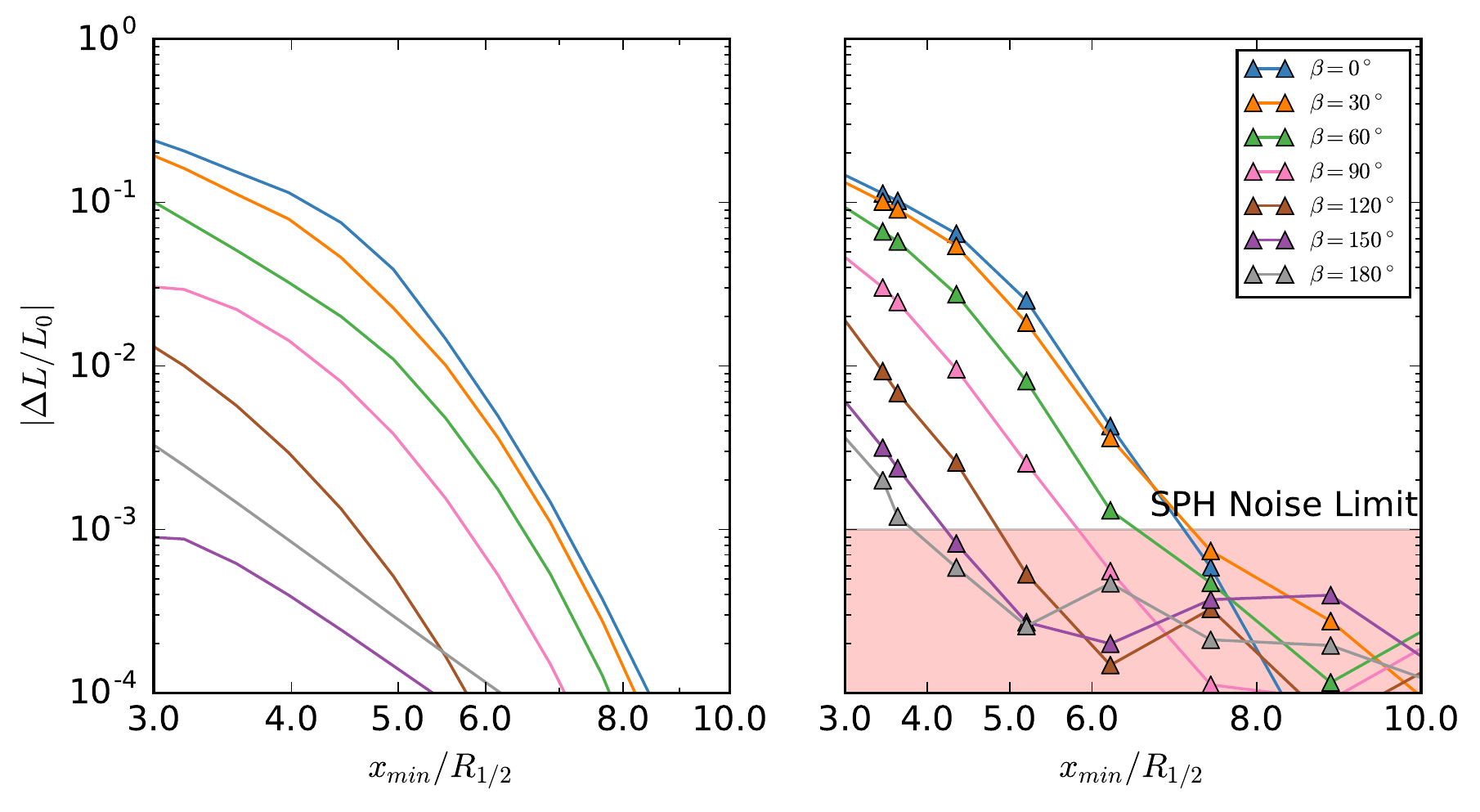}
    \caption{Angular momentum transfer for a disc of half mass radius $R_{1/2}$ around a star of mass $1M_\odot$, perturbed by a star of equal mass for various closest approach distances $x_\mathrm{min}$. The angular momentum vectors between the disc and perturber are offset by various angles $\beta$, while $\alpha=0^\circ$. The left panel is for a disc of test particles, reconstructed from annuli results for analytic power law surface density distribution with $p=1.5$. On the right is the same result for a 3D SPH simulation, with triangle markers representing data points. The region below the empirical noise limit is highlighted in red. }
    \label{fig:plsurf}
\end{figure*}

\begin{figure}
	\includegraphics[width=\columnwidth]{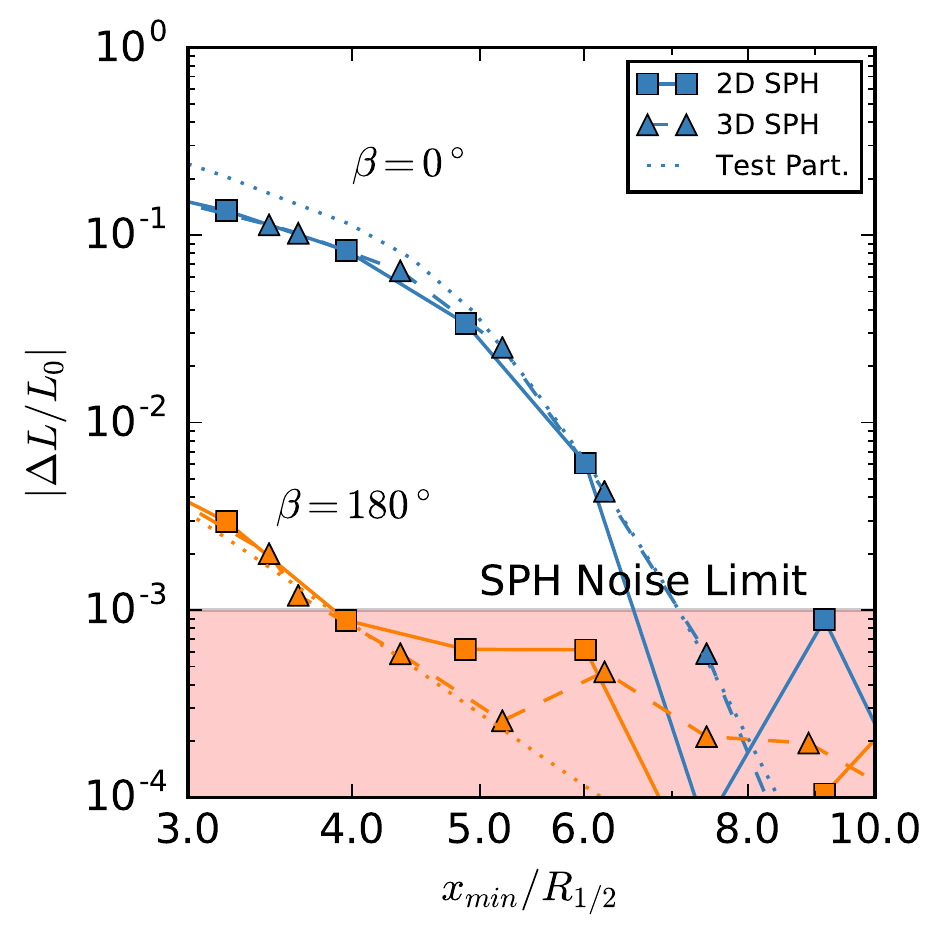}
    \caption{The solid lines (square markers) show the angular momentum transfer for a disc of SPH particles due to a perturber as in Figure \ref{fig:plsurf} in two dimensions, hence with prograde and retrograde trajectories only. The same results are shown for the 3D case (dashed lines, triangle markers) and {for a disc reconstructed from rings of test particles using the results of the \textsc{Mercury} code calculations (dotted lines).  The region in which the SPH calculations become noisy is shaded.} }
    \label{fig:sph_disc2d}
\end{figure}

In our comparisons between the hydrodynamic results and the particle ring ensemble (Figure \ref{fig:plsurf}) the fractional angular momentum in both cases is plotted as a function of the ratio of $x_\mathrm{min}$ to the disc half-mass radius $R_{1/2}$. {This radius is defined at the time of pericentre for an equivalent disc evolving viscously in isolation, although in practice there is little difference between this and the initial value of $R_{1/2}$.} We make this choice of radius here for direct comparison with the results of \citet{Ros14}, and because it is not in general possible to clearly define an outer radius for a viscously evolving disc. The calculations are found to become noisy for $\Delta L/L \lesssim 10^{-3}$. We see good agreement for $x_{\mathrm{min}}/R_{1/2} \gtrsim 4$, and results within a factor of order unity for closer encounters. We note that for our chosen definition of radius (the half-mass radius $R_{1/2}$) with outer radius to inner disc radius ratio $R_\mathrm{out}/R_\mathrm{in}=5$ and $p=3/2$, we have $R_\mathrm{out}\approx1.91 R_{1/2}$, which puts the closest encounters at $x_{\mathrm{min}}/R_{\mathrm{out}} \approx 1.57$. 

The non-zero height of the disc in the 3D case has not significantly altered the results in comparison to the 2D case shown in Figure \ref{fig:sph_disc2d}. The majority of results are dominated by the exponential component, with the exception of the almost retrograde encounters $\beta=150^\circ , \, 180^\circ$ where the $m=1$, $\omega=0$ ILR dominates for all $x_{\mathrm{min}}/r$. We note that in Figure \ref{fig:sph_disc2d} {we have compared the SPH results with the equivalent disc reconstructed out of rings using the results in Section} \ref{sec:ringint} {which do not include viscous and pressure forces. For this comparison we used both the analytic surface density profile and one taken directly from an unperturbed SPH disc at the time of closest approach, but find no significant difference between them. The case shown in Figure} \ref{fig:sph_disc2d} {is for the analytic surface density profile. The resolution and convergence of these results is demonstrated in Appendix} \ref{sec:numconv} {for which no difference is found using $10^6$ particles in 2D above the noise limit. The minor differences compared to the N-body results only play a significant role in strong interactions, where $x_\mathrm{min}/R_{1/2} \lesssim 5$ for a prograde encounter (and angular momentum transfer is non-linear). As these differences are not numerical in origin, they are indicative of hydrodynamical effects not present in the N-body calculations.}

In summary, we have found good agreement across all of our results. The linear calculations detailed in Appendix \ref{sec:reqns} match well with our integration of rings of test particles, and these results in turn agree with full hydrodynamical simulations for interactions such that $|\Delta L/L|> 10^{-3}$. Unfortunately the noise limit of the SPH simulations leaves us unable to test the power law dependence of the angular momentum transfer on $x_{\mathrm{min}}/r$ for distant encounters ($x_\mathrm{min}/r\gtrsim 6$). However these encounters are of little physical significance given their negligible effect on the disc. We conclude that we have robust results for the tidal influence of a gravitationally focused encounter upon a disc.

\section{Discussion}
\label{sec:discuss}

\begin{figure}
	\includegraphics[width=\columnwidth]{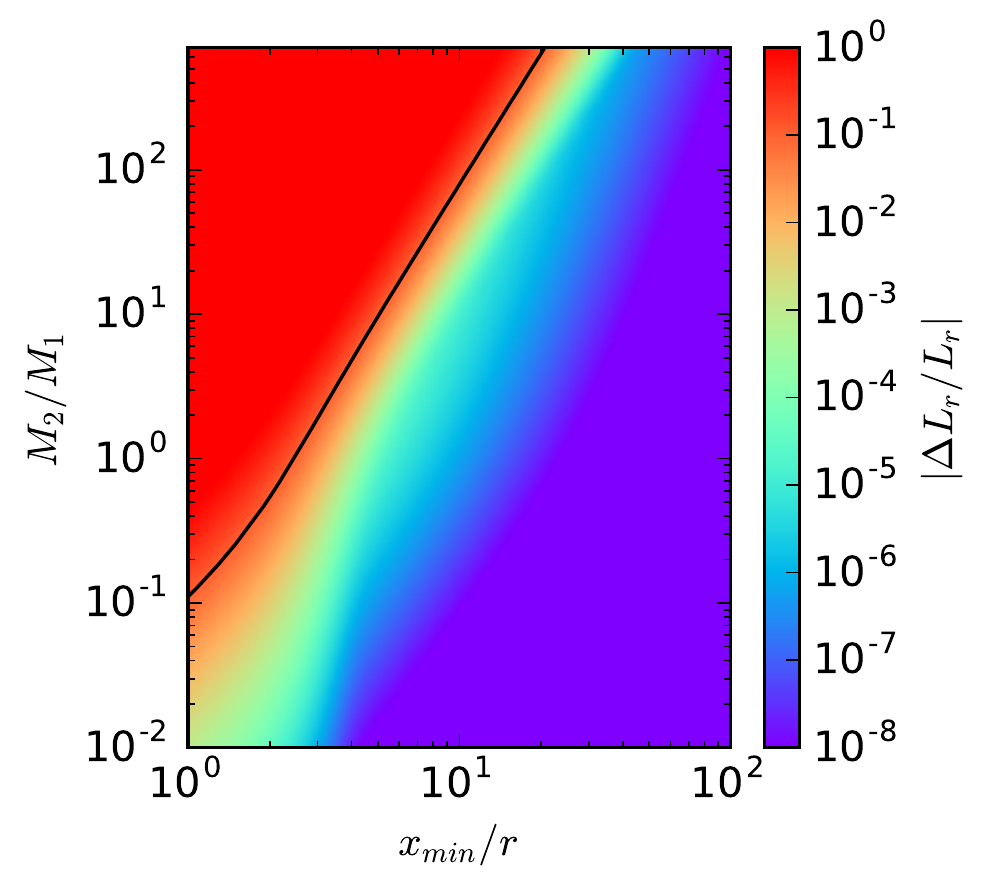}
    \caption{Theoretical angle-averaged parameter space exploration for the fractional angular momentum loss induced for a ring of particles, surrounding a star of mass $M_1$, by a parabolic encounter with a star of mass $M_2$. The contour follows the line at which $\Delta L_\mathrm{r}/L_\mathrm{r} = 0.1$, where the linearised Equations \ref{eq:NLR} and \ref{eq:ILR_an} evaluated here are no longer appropriate.}
    \label{fig:dLmxmin}
\end{figure}

\begin{figure}
	\includegraphics[width=\columnwidth]{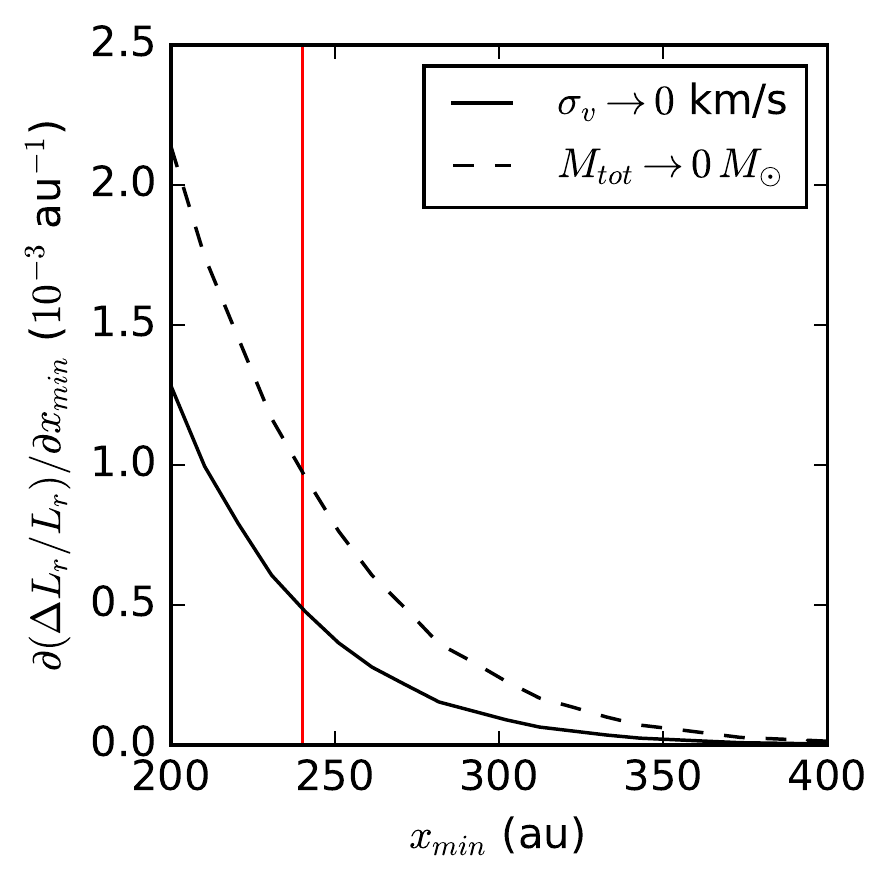}
    \caption{Estimated differential total angular momentum loss for a ring of particles at $R_\mathrm{out} = 100$~au from the host star. Results are shown over a time period such that expected number of encounters such that $x_\mathrm{min}<x_\mathrm{lin}\approx 2.4R_\mathrm{out}$ for each star is unity (see text for details). The value of $x_\mathrm{lin}$ is indicated by the vertical red line. This limit can be generalised for arbitrary mass ratio $M_2/M_1$ by applying the appropriate angular momentum loss threshold, as shown by the black contour in Figure \ref{fig:dLmxmin}. The two most extreme cases for a cluster are shown: the solid line is the limit in which the stellar velocity dispersion is small, and the dashed line is in the limit of an energetic cluster (Equation \ref{eq:diffencrate}).}
    \label{fig:dLdot}
\end{figure}

In order to put these results into the context of stellar encounters in clusters we render the angle averaged results for the fractional angular momentum change per encounter in Figure \ref{fig:dLmxmin} as a function of $x_\mathrm{min}$ and $M_2/M_1$. The black contour represents the point at which the fractional angular momentum loss per encounter is $0.1$ and which we use as the demarcation between linear and non-linear encounters (corresponding to a reduction of the outer disc radius of around $20 \%$). The value of $x_\mathrm{min}$ for which $\Delta L_\mathrm{r}/L_\mathrm{r} =0.1$ we denote $x_\mathrm{lin}$, which is a function of the mass ratio $M_2/M_1$. For closer encounters (i.e. for $x_\mathrm{min} < x_\mathrm{lin}$) the region to the left and above the contour $x_\mathrm{min} = x_\mathrm{lin}$ corresponds to fractional angular momentum loss of order unity. It can immediately be seen from Figure \ref{fig:dLmxmin} that, while more massive perturbers enter the non-linear regime at larger radius, the decline in angular momentum loss in the linear regime is steep for all perturber masses and relates to the exponential decline predicted by equation \ref{eq:ILR_an}.

We illustrate the minor role of encounters in the linear regime by considering the integrated effect of encounters in a stellar population. Following \citet{Bin87}, the differential encounter rate scales as 
\begin{equation}
\label{eq:diffencrate}
\mathrm{d} \Gamma (x_\mathrm{min})  \propto  \left(GM_\mathrm{tot} +  4\sigma_v^2 x_\mathrm{min}\right) \, \mathrm{d} x_\mathrm{min}
\end{equation} for a one dimensional velocity dispersion $\sigma_v$, and total mass  $M_\mathrm{tot}$. We can therefore compute, in a given time $\tau$, the expected fractional
change in angular momentum from encounters in unit interval of $x_\mathrm{min}$ by combining Equation \ref{eq:diffencrate} with the angle averaged change in angular momentum per encounter $ \langle \Delta L_\mathrm{r} /L_\mathrm{r} \rangle$:
\begin{equation}
\label{eq:lossrate}
\frac{\partial (\Delta{L}_\mathrm{r}/L_\mathrm{r})}{\partial x_\mathrm{min}} = \int_0^{\tau}  \left\langle  \frac{\Delta L_\mathrm{r}}{L_\mathrm{r}} \right\rangle  \frac{\partial \Gamma(x_\mathrm{min})}{\partial x_\mathrm{min}} \, \mathrm{d}t
\end{equation} We have assumed in Equation \ref{eq:diffencrate} {that encounters are uncorrelated and make no attempt to include the physical characteristics expected to be present in a cluster, such as internal
substructure} \citep[e.g.][]{Cra13},  {since here we are only interested in the influence of distant encounters relative to close ones and not in the absolute effect of encounters on disc evolution.}

In Figure \ref{fig:dLdot} we depict $\partial (\Delta{L}_\mathrm{r}/L_\mathrm{r})/\partial x_\mathrm{min}$ (Equation \ref{eq:lossrate}) over a time interval such that the expected  number of encounters with $x_\mathrm{min} < x_\mathrm{lin}$ over this time is unity. {We see that the angular momentum change due to the cumulative effect of encounters decreases rapidly with $x_\mathrm{min}$}. Moreover, integration of $\partial(\Delta{L}_\mathrm{r}/L_\mathrm{r})/\partial x_\mathrm{min}$ over $x_\mathrm{min}$ outwards of $x_\mathrm{lin}$ shows that the total fractional angular momentum loss due to encounters in the linear regime over this time interval is $1.7\%$ in the limit of a cold cluster ($\sigma_v \rightarrow 0$), and $4\%$ when the encounter rate (Equation \ref{eq:diffencrate}) is dominated by high velocity encounters. Since the time interval has been chosen so that each star is expect to have experienced one encounter inward of $x_\mathrm{lin}$ over this period, and since such encounters within $x_\mathrm{lin}$ cause an angular momentum reduction of order unity it then follows that the additional effect of encounters outside $x_\mathrm{lin}$ is negligible by comparison. {Note that this time period is dependent on the local physical conditions in the cluster, and depending on the number density and velocity dispersion may be greater or less than the age of a given stellar population.} Although Figure \ref{fig:dLdot} illustrates the situation for equal mass encounters, the exponential fall-off in angular momentum transfer in the linear regime at all masses (Figure \ref{fig:dLmxmin}) means that this minor role for encounters in the linear regime persists for all $M_2/M_1$.

Finally we note that Figure \ref{fig:dLmxmin} shows the region of parameter space for which encounters are at all important, from which a `close-regime' {(i.e. a spatial region in which significant angular momentum loss from the disc occurs)} can be defined for an arbitrary pair of stellar masses. This spatial scale is approximately proportional to $(M_2/M_1)^{1/3}$ which is consistent with the empirical findings of \citet{Bre14} for coplanar, prograde encounters. In order to obtain a general prescription for the effect of close encounters, a full study of the parameter space in this regime is required. Encounters at these large distances for comparatively large $M_2$ are likely to be hyperbolic in hot, high density environments, adding an additional complication. We leave an approach to evaluating the influence of such encounters on a PPD for a future paper.

\section{Conclusions}
\label{sec:conclusion}

We have presented a robust analysis of the angular momentum loss for PPDs during a gravitationally focused encounter. Good agreement is found between the theoretical prescription adapted from \citet{Ost94}, the test particle case explored through N-body simulations, and hydrodynamic simulations for regions of parameter space for which $\Delta L$ is much greater than the numerical noise. 

The angular momentum transfer between disc and perturbing star is dominated by two resonances for non-retrograde encounters; the $m=2$ ILR at small $x_\mathrm{min}/r$ ($\lesssim 6$ for  $M_1=M_2$) and the $m=1$, $\omega=0$ ILR for a secular perturbation at larger encounter distances. For close to retrograde encounters, $150^\circ \lesssim \beta \lesssim 210^\circ$ the $m=2$ ILR contribution is negligible, and hence transfer is dominated everywhere by the $m=1$ ILR. 

We contextualise these results by plotting the angle averaged fractional change in angular momentum per encounter in Figure \ref{fig:dLmxmin} as a function of perturber mass and closest approach distance. Figure \ref{fig:dLmxmin} demonstrates the steep fall off in efficiency of angular momentum transfer in the linear regime for all perturber masses. We show that the total angular momentum loss is always dominated by encounters in the non-linear regime (close encounters at separations less than the black contour shown in Figure \ref{fig:dLmxmin}) and that the angular momentum transfer instead associated with the linear regime is
a small fraction ($<4\%$) of this value.

{In the context of PPD evolution within a stellar cluster, this allows us to conclude that the influence of distant encounters on the disc is negligible, where `distant' is here defined to be any encounter for which the fractional angular momentum loss at the outer edge is $\Delta L/L <0.1$. If a cluster is composed of single mass stars this conclusion is equivalent to the statement that encounters with $x_\mathrm{min}/R_\mathrm{out}>2.4$ can be ignored, despite the increased probability of such an encounter occurring.}

Our theoretical prescription can also be applied to find the upper limit of closest approach distance at which a disc is significantly truncated for a given mass ratio $M_2/M_1$. {Influential encounters in the large $M_2$ regime can occur at comparatively large  $x_\mathrm{min}/R_\mathrm{out}$ and are therefore more likely to be considerably hyperbolic.} We leave a study of the influence of hyperbolic non-linear encounters and their role in stellar clusters for future work.

\section*{Acknowledgements}

We would like to thank the anonymous referee for a considerate report which improved the clarity of this paper. AJW thanks  the  Science  and  Technology  Facilities  Council  (STFC)  for  their  studentship. 
This work has been supported by the DISCSIM project, grant agreement 341137 funded by the European Research Council under ERC-2013-ADG. It has also used the DIRAC Shared Memory Processing system at the University of Cambridge, operated by the COSMOS Project at the Department of Applied Mathematics and Theoretical Physics on behalf of the STFC DiRAC HPC Facility (www.dirac.ac.uk). This equipment was funded by BIS National E-infrastructure capital grant ST/J005673/1, STFC capital grant ST/H008586/1, and STFC DiRAC Operations grant ST/K00333X/1. DiRAC is part of the National E-Infrastructure.




\bibliographystyle{mnras}
\bibliography{truncation} 


\onecolumn
\appendix
\section{Linear Angular Momentum Transport for Rings}
\label{sec:reqns}
In order to derive results for a given ring at radius $r$ within a disc, we adapt the results of \citet{Ost94}, and henceforth equation numbers in brackets are in reference to that paper. Before we discuss the individual contributions, some consideration is given to particular Equations and notation within that extensive study. We first briefly review the relevant vectors. The vector between the star with a disc and a given fluid element is defined to be
	$$
	\bm{r} =\bm{r}_0(t) + \bm{r}_1
	$$ where $\bm{r}_0$ is the unperturbed position vector and $\bm{r}_1$ is the perturbation from this vector induced by the force from the secondary star. This can be expanded into the form 
	$$
	\bm{r}_1\equiv r_1 \hat{\bm{r}} + \phi_1 r_0 \hat{\bm{\phi}} + z_1 \hat{\bm{z}}
	$$	where subscript $0$ denotes an unperturbed co-ordinate value, and $1$ the corresponding perturbed value. Note that $z_0=0$ in the chosen coordinate system. 
	The external force per disc fluid element mass exerted by the perturbing star with mass $M_2$, separated by vector $\bm{x}$ from the host star, can be expanded in terms of spherical harmonics:
	\begin{equation}
	\label{eq:fext}
	\bm{f}_{\mathrm{ext}} = GM_2 \mathlarger{\sum}_{l=2}^\infty \mathlarger{\sum}_{m=-l}^l \frac{4\pi}{2l+1} \bm{\nabla} \left[|\bm{r}'|^l Y_l^{m*}(\bm{r}')\right]\frac{Y_l^m(\bm{x})}{|\bm{x}|^{l+1}} 
	\end{equation} where $\bm{r}'$ is the distance between the perturbed fluid particle and the center of mass of the system
	$$
	\bm{r}' = \bm{r} - \frac{M_{\mathrm{disc}}}{M_1+M_{\mathrm{disc}}} \bm{r}_{\mathrm{disc}} \approx \bm{r}
	$$ assuming that the mass of this disc $M_{\mathrm{disc}}$ is negligible. 
	
	While working with spherical harmonics in this context it becomes necessary to define a quantity denoted in \citet{Ost94} as $Y_l^m(0)$. This quantity is used to represent the polar part of the spherical harmonic evaluated at $\bm{r}_0$, which is $Y_l^m(\pi/2,0)$ given that the disc is fixed in the equatorial plane. For Equation [2.15] and [2.16] the azimuthal component is cancelled when the Laplace transformation is applied to Equations [2.10] and [2.11]. The integral over $\phi_0$ is then simply a factor $2\pi$.
		
	To calculate the angular momentum change in the disc, it is noted that $\mathrm{d}\bm{L} \approx \mathrm{d}m \, \bm{r}\times \mathrm{d}\bm{v} =  \mathrm{d}m \, \bm{r} \times \bm{f}_{\mathrm{ext}} \, \mathrm{d} t$,  so we have
	\begin{equation}
	\label{eq:dL_vect}
	\Delta \bm{L} = GM_2 \int_{R_{\mathrm{in}}}^{R_{\mathrm{out}}} \mathrm{d}r_0 \, r_0 \Sigma_0(r_0) \, \int_0^{2\pi} \mathrm{d}\phi_0 \int_{-\infty}^{\infty} \mathrm{d} t \, \bm{r} \times \bm{f}_{\mathrm{ext}}
	\end{equation} From  Equation \ref{eq:linapprox}, we can approximate the total change in angular momentum by the $z$-component of Equation \ref{eq:dL_vect}. Although \citet{Ost94} finds a component proportional to $Y^{m+1}_l(0)$ and the Laplace-transformed $z$-coordinate, we show  briefly that this term is not present in the vertical projection of the angular momentum transfer. This is important because it is the term which eventually leads to a vertical resonance contribution which dominates at large $x_{\mathrm{min}}/r$ in the calculations of \citet{Ost94}.
	Substituting Equation \ref{eq:fext} into Equation \ref{eq:dL_vect} and expanding by the product rule, it is immediately clear that any $Y^{m+1}_l$ terms must come from the expression $\bm{r} \times \bm{\nabla} Y_l^{m*}(\bm{r})$. This is easier to evaluate in spherical coordinates $\rho$, $\theta$, $\phi$, in which case $\bm{\nabla} Y_l^{m*}(\bm{r})$ has $\hat{\bm{\theta}}$ and  $\hat{\bm{\phi}}$ components only, which we denote $\nabla_\theta Y_l^{m*}(\bm{r})$ and $\nabla_\phi Y_l^{m*}(\bm{r})$. Now taking the cross product with $r$, we find
	$$
	\bm{r} \times \bm{\nabla} Y_l^{m*}(\bm{r}) = \left[\theta \nabla_\phi Y_l^{m*}(\bm{r}) -\phi \nabla_\theta Y_l^{m*}(\bm{r})\right] \hat{\bm{\rho}} - \rho \nabla_\phi Y_l^{m*}(\bm{r}) \hat{\bm{\theta}} + \rho \nabla_\theta Y_l^{m*}(\bm{r})\hat{\bm{\phi}} 
	$$Changing to cylindrical unit vectors, the $z$-component of this product is 
	$$
	\hat{\bm{z}} \cdot \left[\bm{r} \times \bm{\nabla} Y_l^{m*}(\bm{r}) \right] =  \left[\theta \nabla_\phi Y_l^{m*}(\bm{r}) -\phi \nabla_\theta Y_l^{m*}(\bm{r})\right] \cos \theta + \rho \nabla_\phi Y_l^{m*}(\bm{r}) \sin \theta \approx \frac{\rho (r_0+r_1)\nabla_\phi Y_l^{m*}(\bm{r})}{\sqrt{(r_0+r_1)^2+z_1^2}}
	$$ The last equality is true because $\cos \theta = z_1/\sqrt{(r_0+r_1)^2+z_1^2}$, which is second order compared with $\sin \theta $. This final step relies on the assumption that the polar component of the derivative is of the same order or smaller than the azimuthal derivative, but this is fine. In fact the final value for this element simplifies to
	 $$
	 \hat{\bm{z}} \cdot \left[\bm{r} \times \bm{\nabla} Y_l^{m*}(\bm{r}) \right] = -i m Y_l^{m*}(\bm{r})
	 $$Hence, the contribution of the vertical resonance is not dominant at large closest approach distances, and we find instead an ILR dominates angular momentum transfer in all regimes.
	 
	 We now jump to the derived expression for the Lindblad resonances, which turn out to be the dominant contributions to the angular momentum loss in the disc. Equation [2.43] for angular momentum is 
	                                                                                                                                                                                                                                                                                                                                                                                                                                                                                                                                                                                                                                                                                                                                                                                                                                                                                                                                                                                                                                                                                                                                                                                                                                                                                                                                                                                                                                                                                                                                                                                                                                                                                                                                                                                                                                                                                                                                                                                                                                                                                                                                                                              \begin{equation}
	                                                                                                                                                                                                                                                                                                                                                                                                                                                                                                                                                                                                                                                                                                                                                                                                                                                                                                                                                                                                                                                                                                                                                                                                                                                                                                                                                                                                                                                                                                                                                                                                                                                                                                                                                                                                                                                                                                                                                                                                                                                                                                                                                                              \label{eq:ILROLR}
\Delta L^{\mathrm{ILR/OLR}} \! \! \! = \pm \mathlarger{\sum}_{m=0}^\infty \frac{1}{1+ \delta_{m0}}  \int_{\omega_{\mathrm{min}}}^{\omega_{\mathrm{max}}}  \! \! \! \! \! \!\mathrm{d}\omega \,\frac{ m \pi \Sigma_0(r_{\mathrm{L}})}{{r_{\mathrm{L}} \kappa(r_{\mathrm{L}}) \partial(m\Omega\mp\kappa)/\partial r_{\mathrm{L}}} } 
\left| GM_2 \mathlarger{\sum}^{\infty}_{l \geq |m|,2}  \frac{4\pi}{2l+1}  r_{\mathrm{L}}^l \left(l\pm \frac{2m\Omega(r_{\mathrm{L}})}{\kappa(r_{\mathrm{L}})}\right) Y_l^m(0)  \mathlarger{\int}^{\infty}_{-\infty}  \! \! \mathrm{d}t \, \frac{Y_l^m(\bm{x})}{|\bm{x}|^{l+1}} e^{-i\omega t}\right|^2                                                                                                                                                                                                                                                                                                                                                                                                                                                                                                                                                                                                                                                                                                                                                                                                                                                                                                                                                                                                                                                                                                                                                                                                                                                                                                                                                                                                                                                                                                                                                                                                                                                                                                                                                                                                                                                                                                                                                                                                                                                                                                                                                                              \end{equation} 
 where $r_L$ is the radius defined such that $\omega = m\Omega(r_L) \mp \kappa(r_L)$ are satisfied for ILR/OLR respectively, where $\kappa$ is the epicyclic frequency. The limits $\omega_{\mathrm{min/max}}$ are defined similarly for the maximum and minimum disc radii. As discussed in Section \ref{sec:lineqs}, the epicyclic and Keplerian frequencies coincide when the disc mass is negligible. In this case, $\omega$ apparently vanishes for the $m=1$ contribution; the forcing frequency is small enough such that, although the radius of exact resonance is not well defined, a large range of radii are in a state of near resonance (and exact resonances do not exist). Formally the outer radius for which this is true is described in terms of a fiducial `wave radius' $r_w$ \citep{Ost92}. However, for a low-mass disc, the effect of the near resonance are nearly the same as the $m=1$ contribution to the exact resonance in Equation \ref{eq:ILROLR} and setting $\omega=0$. This is physically equivalent to a secular perturbation, wherein the external trajectory is replaced by a ring of mass per unit length proportional to the inverse of the velocity at each point. We then have $\mathrm{d}\omega \cdot \left[\partial(\Omega - \kappa)/ \partial r_{\mathrm{L}} \right]^{-1} \approx \mathrm{d}r$, from which comes Equation [2.50]. The integration limits are swapped, and the value of $m$ is negated. The secular resonance acts as an OLR since it propagates outwards. We are interested in the angular momentum transferred to a ring within the disc. This is obtained by treating the surface density distribution as a delta function $\Sigma_0(r) \rightarrow \delta(r-r_{\mathrm{r}})$, where the subscript `r' denotes the ring quantity. Dividing through by the total angular momentum
 $$
 L = 2\pi \int_{R_{\mathrm{in}}}^{R_{\mathrm{out}}} r^3\Sigma_0(r) \Omega(r) \, \mathrm{d} r 
 $$ where $\Sigma_0(r)$ is again treated as a delta function gives $\Delta L_{\mathrm{r}}/L_{\mathrm{r}}$.
 
 The final stage is to parameterise the trajectory of the perturbing star  $\bm{x}$ relative to the central star, and the time $t$ in terms of an angular coordinate $\psi$, phase with respect to pericentre. The appropriate transformation for $\tau = \tan(\psi/2)$ is
  $$
    x(\tau)  = x_{\mathrm{min}} (1+\tau^2) \, ; \quad t = \left( \frac{2x_{\mathrm{min}}^3}{GM_{\mathrm{tot}}} \right)^{1/2} \left( \tau + \tau^3\right)
    $$ so that the time integral is over $\psi$ between $-\pi$ and $\pi$. Hence for the $m=1$, $\omega=0$ case the contribution of the ILR to angular momentum loss from a ring at radius $r$ within the disc is
\begin{equation}
   	\label{eq:NLR}
	\left.\frac{\Delta L_{\mathrm{r}}}{L_{\mathrm{r}}}\right|_{m=1, \,\omega=0}^{\mathrm{ILR}} \! \! \! \! \! \! \! \! \!  = - \frac{M_2^2}{4M_1M_{\mathrm{tot}}} \frac{x_{\mathrm{min}}}{r}\left|\mathlarger{\sum}^{\infty}_{l \geq 2}  \frac{4\pi(l+2)}{2l+1} \left( \frac{x_{\mathrm{min}}}{r}\right)^{-l} Y_l^1(0) \mathlarger{\int}^{\pi}_{-\pi} \! \! \mathrm{d}\psi \, \cos^{2l-2}(\psi/2) Y^{1}_{l} (\bm{x}) \right|^2
	\end{equation} where it is understood that the spherical harmonic as a function of the separation can be rotated in the axes described in Section \ref{sec:ringmeth} such that 
	
	$$
	Y^{m}_{l} (\bm{x}) = \mathlarger{\sum}^{l}_{m'=-l} Y^{m'}_{l} (\pi/2, \psi ) d^{l} _{m' m} (\beta) e^{-im\alpha}
	$$ where $d^{l} _{m' m}$ is a Wigner-d matrix.
	
	The contributions of the exact resonances can be found in a similar way to be 
\begin{equation}
	\begin{split}
   	\label{eq:ILR}
	&\left.\frac{\Delta L_{\mathrm{r}}}{L_{\mathrm{r}}}\right|^{\mathrm{ILR}} \! \! \! \!  = -\mathlarger{\sum}_{m=2}^\infty  \frac{ m M_2^2}{2M_1 M_{\mathrm{tot}}} \frac{x_{\mathrm{min}}}{r} \left|\mathlarger{\sum}^{\infty}_{l = m}  \frac{4\pi(l-2m)}{2l+1} \left( \frac{x_{\mathrm{min}}}{r}\right)^{-l} Y_l^m(0) \mathlarger{\int}^{\pi}_{-\pi} \! \! \mathrm{d}\psi \, \cos^{2l-2}(\psi/2) Y^{m}_{l} (\bm{x}) \exp \left[-i 2^{3/2}y \frac{\tan (\psi/2)}{\cos^2(\psi/2)}  \right] \right|^2
	\end{split}
	\end{equation} where
	$$
	y = (m-1)\left( \frac{M_1}{M_\mathrm{tot}} \right)^{1/2}  \left( \frac{x_\mathrm{min}}{r} \right)^{3/2}
	$$ Equation \ref{eq:NLR} has the clear properties that it scales as $(x_{\mathrm{min}}/r)^{1-2l}$, and is therefore dominated by the non-zero components with lowest values of $l$. As $Y_2^1(0)=0$, this is the $l=3$ term. While Equation \ref{eq:ILR} can simply be calculated numerically, in order to write the exact resonance contributions in a helpful form an approximation for the integral over $\psi$ (for $\omega \neq 0$) is needed. The approach for this is provided in \citet{Ost94}, and we obtain
	\begin{equation}
	\label{eq:ILR_an}
	  \left.\frac{\Delta L_{\mathrm{r}}}{L_{\mathrm{r}}}\right|^{\mathrm{ILR}} \! \! \! = -\mathlarger{\sum}_{m=2}^\infty   \frac{m \pi M_2^2}{2M_1 M_\mathrm{tot}}  \frac{x_\mathrm{min}}{r}\exp \left[ -\frac{2^{5/2}}{3} y \right] 
	\left|\mathlarger{\sum}^{\infty}_{l = m}  \frac{4\pi(l-2m)}{2l+1} \frac{2^{3l/2+1/4}}{(2l-1)!!} y^{(2l-1)/2} \left( \frac{x_\mathrm{min}}{r}\right)^{-l} Y_l^m(0) Y_l^l(\bm{x}_\mathrm{min}) \right|^2
	\end{equation} This term is dominated by lower $m$ values, and is thus referred to as the $m=2$ ILR component. The total angular momentum lost in a close encounter can be approximated by the sum of Equations \ref{eq:NLR} and \ref{eq:ILR_an}.
	
\twocolumn
\section{Numerical Convergence Tests}
\label{sec:numconv}

\begin{figure}
	\includegraphics[width=\columnwidth]{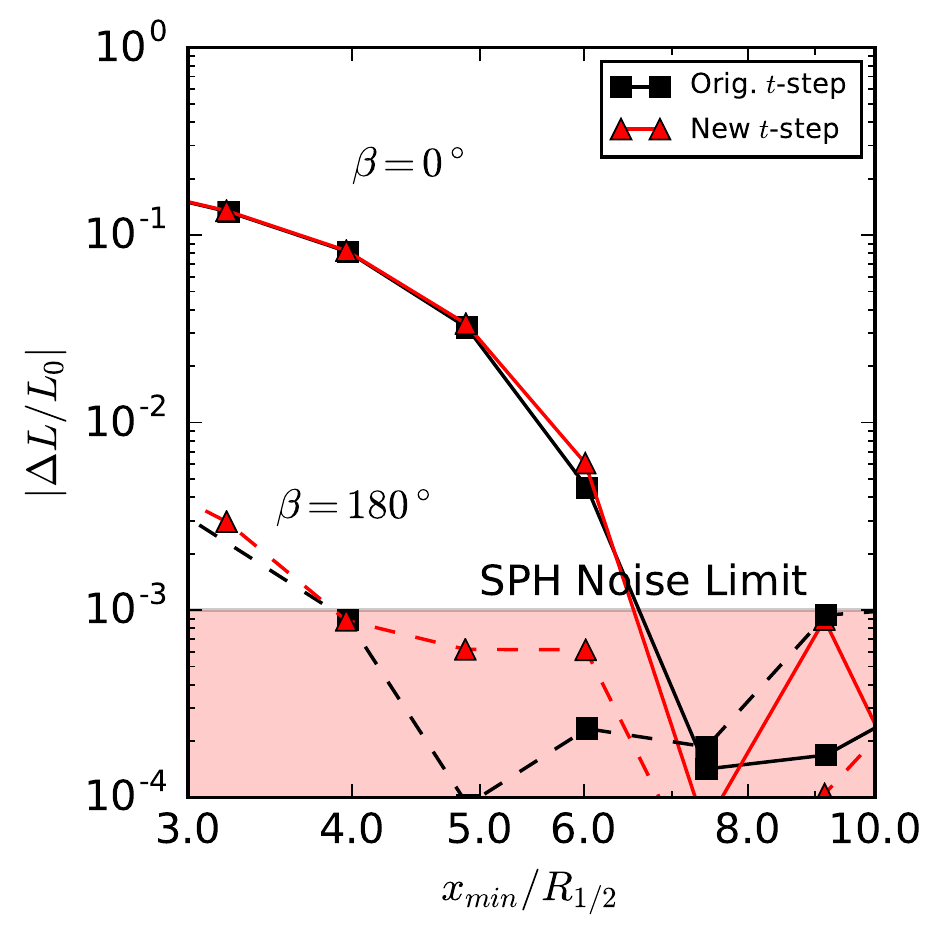}
    \caption{Time-step test for 2D SPH calculation results. The new results (red) use a smaller a time-step which is reduced by a factor three. This is compared to our original results (black) in the case where the encounter is prograde (solid) and retrograde (dashed).  }
    \label{fig:tsteptest}
\end{figure}

\begin{figure}
	\includegraphics[width=\columnwidth]{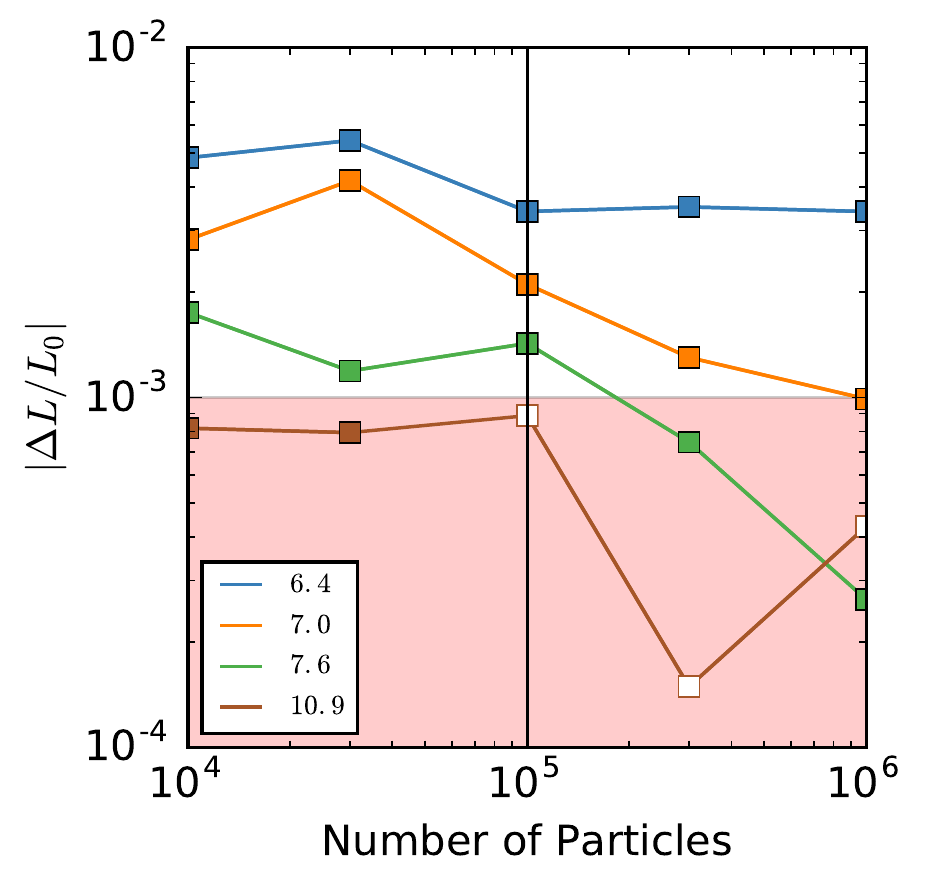}
    \caption{Change in angular momentum for the prograde case in 2D. Numbers in the legend represent the value of $x_\mathrm{min}/R_{1/2}$. Results for which $\Delta L/L>0$ are shown as empty squares, while filled squared represent angular momentum loss. The vertical line is placed at $10^5$ particles, which is the resolution of the 2D SPH results presented in Section \ref{sec:sphresults}. The noise limit found for $10^5$ particles is shaded.}
    \label{fig:restest}
\end{figure}
\begin{figure}
	\includegraphics[width=\columnwidth]{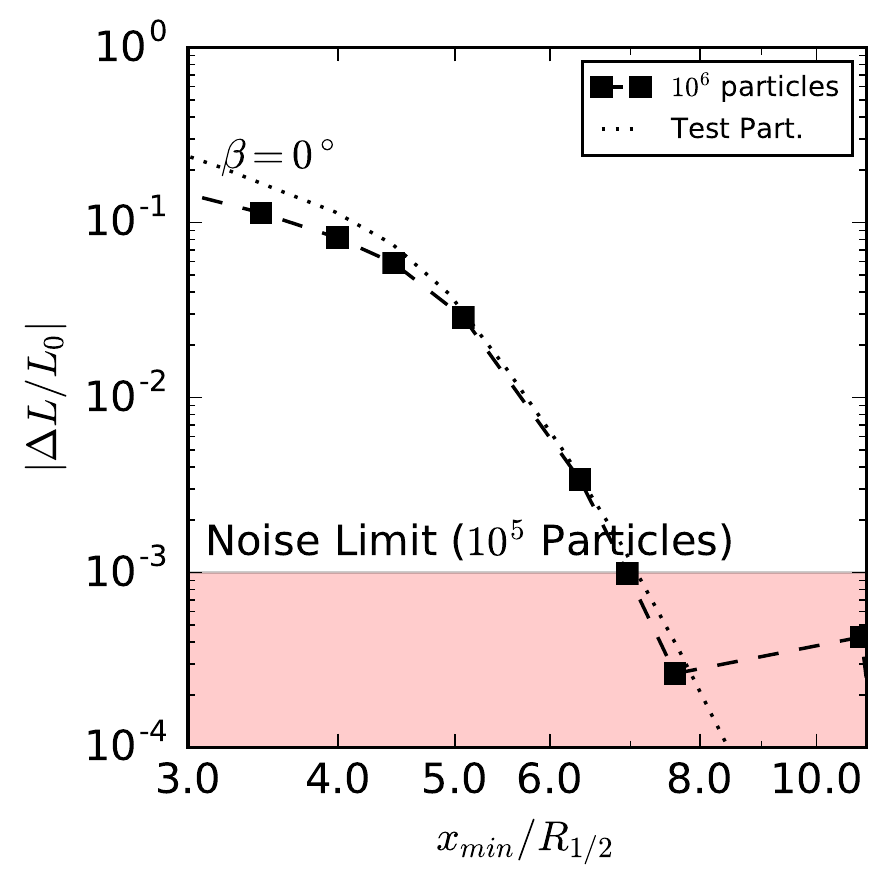}
    \caption{ Angular momentum change calculations using SPH in 2D for a prograde encounter with $10^6$ particles. The test particle reconstructed disc case is plotted for comparison, as in Figure \ref{fig:sph_disc2d}. The shaded region represents the assumed noise limit in the $10^5$ particle case.  }
    \label{fig:incres}
\end{figure}

 To ensure that the results of our SPH calculations using \textsc{Gandalf} are numerically converged, we show the equivalent results in the 2D disc case for different particle resolutions and alternate time-step criteria.

In our discussion of SPH results, we compare simulations performed with a Leapfrog time integration method to those of \textsc{Mercury}'s built in BS integrator. While the latter technically has much greater accuracy, we show that the Leapfrog integrator is accurate enough for the range of parameter space we are interested in. To test the time-step criteria, we reduce the code time-steps such that temporal resolution is improved by a factor of $3$. The results of this procedure are shown in Figure \ref{fig:tsteptest}. No significant difference is found between the two sets of results above the $|\Delta L/L| \sim 10^{-3}$ noise limit, which also remains unchanged.

To test the resolution dependence of our results we run the 2D prograde simulations with $10^4$, $3\times 10^4$, $10^5$, $3\times 10^5$ and $10^6$ particles. We plot angular momentum change for $x_\mathrm{min}/R_\mathrm{1/2}>6$ as a function of resolution in Figure \ref{fig:restest}. Results with $x_\mathrm{min}/R_\mathrm{1/2}\lesssim 6$ are converged down to particles numbers $\sim 10^4$. We find that the noise limit is relatively insensitive to particle number. Previous work has found SPH accuracy to be limited by kernel choice (number of particle neighbours) and the sound speed within the disc. Given our choice of a cubic kernel it is possible that noise limit might be improved by applying, for example, a higher order kernel \citep{Boo15, Ross15}. However, given that the required accuracy has been achieved in the region of interest, we do not pursue this possibility further here. 

We additionally plot the $10^6$ SPH particle results compared to the test particle disc reconstruction in Figure \ref{fig:incres}. Good agreement is found over the the same range as the $10^5$ case, with some marginal improvement at $x_\mathrm{min}/R_\mathrm{out} \sim 8$. This confirms our suggestion in Section \ref{sec:sphresults} that the disc hydrodynamics slightly alter angular momentum transfer in the close regime $x_\mathrm{min}/R_{1/2}<5$. 

We conclude that the SPH results that we present in Section \ref{sec:sphresults} are not resolution dependent in the high $|\Delta L/L| >10^{-3}$ regime, and are not altered by improving the time-step criteria. This justifies our comparison between the \textsc{Mercury} particle ring calculations and the SPH disc calculations using \textsc{Gandalf}.

\bsp	
\label{lastpage}
\end{document}